\documentclass[12pt]{article}
\usepackage{amsfonts, latexsym}
\usepackage{amsmath} 
\usepackage{amsthm}
\usepackage{bbm}
\usepackage{array}
\begin{document}

\begin{center}
{\Large \bf Gauge Invariant Formulation and Bosonisation of the Schwinger Model}
\ \\ \ \\
{\large J. Kijowski} \\ 
{\normalsize Center for Theor. Phys., Polish Academy of Sciences \\
al. Lotnik\'ow 32/46, 02 - 668 Warsaw, Poland} \\ 
{\large G. Rudolph, M. Rudolph} \\
{\normalsize Institut f\"ur Theoretische Physik, Univ. Leipzig \\
Augustusplatz 10/11, 04109 Leipzig, Germany}
\end{center}

\begin{abstract}
The functional integral of the massless Schwinger model in $(1+1)$ dimensions
is reduced to an integral in terms of local gauge invariant quantities.
It turns out that this approach leads to a natural bosonisation scheme,
yielding, in particular the famous ``bosonisation rule'' and giving some
deeper insight into the nature of the bosonisation phenomenon.
As an application, the chiral anomaly is calculated within this formulation.

keywords: schwinger model; gauge invariants; bosonisation; chiral anomaly 

PACS: 11.15.-q; 11.10.Kk; 11.25.Sq; 11.30.Rd
\end{abstract}

\section{Introduction}

In recent years there has been some effort to understand the classical
bosonisation results for models in (1+1) dimensions, see \cite{C}, \cite{Fr}
and further references therein, in terms of functional integral techniques, see
\cite{DN1,DN2,DN3}, \cite{SW} and further references therein. In this Letter
we apply our programme of formulating gauge theories in terms of
local gauge invariant quantities, see \cite{KR}, \cite{KRR1} and \cite{KRR2},
to the massless Schwinger model in 2 dimensions, see \cite{Sch}.
We show that it leads to a natural explanation of the bosonisation phenomenon,
including mass generation. We also demonstrate that our formulation provides
us with the possibility to calculate the chiral anomaly in a straightforward,
simple way. We stress that this is in contrast to calculations within the
``ordinary'' formulation, where one either uses perturbative techniques or
topologically motivated regularizations (heat kernel, zeta--function), see
e.g. \cite{B}.

Our approach is based upon the following ideas: First one has to analyze the
algebra of Grassmann-algebra-valued invariants, which one can build from the
gauge field and the anticommuting matter fields. For a systematic study of
this aspect for the (most complicated) case of QCD we refer to \cite{KRR3}.
Typically, there occurs a number of
relations between invariants, which on the level of the algebra cannot be
``solved'', because dividing by Grassmann-algebra-valued quantities is in
general not well defined. Next one has to express the Lagrangian, or at least
the Lagrangian multiplied by some nonvanishing element of the above algebra,
in terms of invariants. The main point of our strategy consists in reducing
the original functional integral measure to a measure in terms of invariants,
for details we refer to \cite{KRR1} and \cite{KRR2}. In particular, under the
functional integral we are able to solve the above-mentioned relations between
invariants, leading to theories with the correct number of effective fields.
Along these lines we have found, both for QED and QCD, a functional integral
formulation in terms of effective bosonic fields.

The paper is organized as follows: In Section 2 we introduce some basic
notations and give the definition of gauge invariants, built from the gauge 
potential and the (anticommuting) spinor fields. Some algebraic 
relations among these invariants, necessary for our purposes, are written down.
The main result of Section 3 is the functional integral completely rewritten
in terms of local (bosonic) gauge invariants. Finally, in Section 4 we derive
the standard bosonisation rule and calculate the chiral anomaly.

\setcounter{equation}{0}
\section{Basic Notations and Gauge Invariants}

A field configuration of the Schwinger model \cite{Sch} is
given by a $U(1)$-gauge potential $A_{\mu}$ and a spinor field $\psi^K$.  
In what follows $\mu,\nu,\ldots = 0,1$ denote space--time indices
and $K,L,\ldots=1,2$ spinor indices. The spinor field is represented by
\begin{equation}
\label{2.0}
\psi^K= \left( \begin{array}{c}
	\phi \\
	\varphi 
	\end{array} \right), 
\hspace*{2cm}
\psi_K^{\dagger} = \left( \phi^*, \varphi^* \right),
\end{equation}        
where the star denotes complex conjugation.
The 4 elements $(\psi^K, \psi_K^*)$ anticommute and (pointwise)
generate a 16-dimensional Grassmann-algebra.

The Dirac--matrices in Minkowski space with metric
$\eta^{\mu \nu} = {\rm diag}(1,-1)$ are taken as
\begin{equation}
\label{2.1}
\gamma^0 = \left( \begin{array}{cc} 0 & 1 \\
				    1 & 0
		  \end{array} \right),			 
\hspace*{0.5cm}
\gamma^1 = \left( \begin{array}{cc} 0 & -1 \\
				    1 &  0
		  \end{array} \right),			 
\hspace*{0.5cm}
\gamma^5 = \left( \begin{array}{cc} 1 &  0 \\
				    0 & -1
		  \end{array} \right).
\end{equation}
The completely antisymmetric symbol in 2 dimensions is denoted by
$\epsilon_{\mu \nu}$, with $\epsilon_{01} = 1$.

The starting point for quantum theory is the following (formal) functional
integral:
\begin{equation}
\label{2.2}
Z = \int \prod_x {\rm d}A \, {\rm d}\psi \, {\rm d}{\bar \psi} \,
    e^{{\rm i} \, \int {\rm d}^2x \, {\cal L}},
\end{equation}		 
with the Lagrangian
\begin{equation}
{\cal L} = {\cal L}_{gauge} + {\cal L}_{mat} =
- \, \tfrac{1}{4} \, F_{\mu \nu} F^{\mu \nu} - \,
{\rm Im} \left\{\overline{\psi_K} \, {(\gamma^{\mu})^K}_L \,
\left( D_{\mu} \psi^L \right) \right\}. \label{2.3}
\end{equation}
Here $ F_{\mu \nu} = \partial_{\mu} A_{\nu} - \partial_{\nu} A_{\mu}$ denotes 
the field strength,
$D_{\mu} \psi^K = \partial_{\mu} \psi^K + i e \, A_{\mu} \psi^K$ the covariant 
derivative and $\overline{\psi_L} := \psi_K^{\dagger} \, {(\gamma^0)^K}_L
= \left( \varphi^*, \phi^* \right).$		

Now we start to apply our approach to this model. For this purpose we
define the following set of gauge invariant Grassmann-algebra valued
quantities:
\begin{eqnarray}
L & := & \varphi^* \, \varphi, \qquad{} 
R := \phi^* \, \phi, \qquad{} 
H := \phi^* \, \varphi, \label{2.4} \\
B_{\mu} & := & \tfrac{1}{2} \, {\rm Im} \left\{ H^* \, 
	       \left(\phi^* \, (D_{\mu} \varphi) + \varphi \, (D_{\mu} \phi^*)
	       \right) \right\}. \label{2.5}
\end{eqnarray}
Obviously, $L$, $R$ and $H$ are (pointwise) elements of rank two, obeying the
following identity:
\begin{equation}
L \, R = - |H|^2. \label{2.6} 
\end{equation}
Moreover, $|H|^2 = H H^*$ is a nonvanishing element of (pointwise)
maximal rank. The vector current is defined by
\begin{equation}
\label{2.7}
J^{\mu} := \overline{\psi_K} \, {(\gamma^{\mu})^K}_L \, \psi^L
	 = \left( \begin{array}{c} R+L \\ R-L \end{array} \right). 
\end{equation}	  
In 2 dimensions we have the relation
$J^{\mu}_5 = \epsilon^{\mu \nu} J_{\nu}$.
Thus, we get for the axial current
\begin{equation}
\label{2.8}
J^{\mu}_5 = \left( \begin{array}{c} R-L \\ R+L \end{array} \right).
\end{equation}
Finally, for $B_{\mu}$ we have
\begin{equation}
\label{2.9}
B_{\mu} = e \, |H|^2 \, A_{\mu} 
	  + \tfrac{1}{4 {\rm i}} \, \left\{ 
	    \varphi^* \phi^* \, \partial_{\mu} (\varphi \phi) 
	    + \varphi \phi \, \partial_{\mu} (\phi^* \varphi^*) \right\}.
\end{equation}

\setcounter{equation}{0}
\section{Functional Integral}

In order to be able to calculate expectation values of physical quantities one 
usually applies the Faddeev--Popov gauge fixing procedure, see \cite{FP} and
\cite{FS}, to the functional integral (\ref{2.2}).
Here we use a completely different procedure, proposed in \cite{KRR1} and
\cite{KRR2}, which consists in reducing (\ref{2.2}) to an integral in terms 
of the above defined gauge invariants. 

For that purpose we introduce the following auxiliary gauge invariants:
\begin{equation}
\label{3.1}
C_{\mu} := \phi^* \, (D_{\mu} \varphi), \qquad{}
E_{\mu} := \varphi^* \, (D_{\mu} \varphi), \qquad{}
F_{\mu} := \phi^* \, (D_{\mu} \phi),
\end{equation}
which are subject to the relations:
\begin{eqnarray}
H^* C_{\mu} & = & {\rm i} \, B_{\mu} + \tfrac{1}{2} \, H^* (\partial_{\mu} H)
		  - \, \tfrac{1}{4} \, R (\partial_{\mu} L)
		  + \, \tfrac{1}{4} \, L (\partial_{\mu} R), \label{3.2} \\
R E_{\mu} & = & - {\rm i} \, B_{\mu} - \tfrac{1}{2} \, H^* (\partial_{\mu} H)
		  + \, \tfrac{1}{4} \, R (\partial_{\mu} L)
		  - \, \tfrac{1}{4} \, L (\partial_{\mu} R), \label{3.3} \\
L F_{\mu} & = & - {\rm i} \, B_{\mu} + \tfrac{1}{2} \, H^* (\partial_{\mu} H)
		  - \, \tfrac{1}{4} \, R (\partial_{\mu} L)
		  + \, \tfrac{1}{4} \, L (\partial_{\mu} R)
		  + \, \tfrac{1}{2} \partial_{\mu}( LR ). \label{3.4} 
\end{eqnarray}
One easily proves the following identities:
\begin{eqnarray}
e \, |H|^2 F_{\mu \nu}
 & = & - {\rm i} \, \partial_{[\mu} ( |H|^2 C_{\nu]} )
       + {\rm i} \, C_{[\mu}^* C_{\nu]} 
       - {\rm i} \, (\partial_{[\mu} R) \, E_{\nu]}, \label{3.6} \\
{\cal L}_{mat} 
 & = & - \tfrac{1}{2} \,
       {\rm tr}\left( \gamma^0 \gamma^{\mu} \right)
       \left\{ (\eta_{\mu \nu}+\epsilon_{\mu \nu}) \, {\rm Im} F^{\nu}
       + (\eta_{\mu \nu}-\epsilon_{\mu \nu}) \, {\rm Im} E^{\nu} \right\} 
\nonumber\\
& \equiv &  - {\rm Im}\left\{ E_0 + F_0 - E_1 + F_1 \right\}  \, .
\label{3.7}
\end{eqnarray}
Observe, that (\ref{3.6}) is an identity on the level of maximal 
rank in the Grassmann--algebra. Dividing by a non--vanishing element of maximal 
rank is a well--defined operation giving a c--number. Thus, we can express the
Lagrangian (\ref{2.3}) completely in terms of gauge invariants.

With the identities (\ref{3.2})--(\ref{3.7}) we are now able to reduce the 
functional integral (\ref{2.2}) to an integral over gauge invariants. For that
purpose we make use of the following notion of the $\delta$--distribution on
superspace (see \cite{Ber2})
\begin{equation}
\delta(u - U) 
 = \int {\rm d} \xi \, e^{2 \pi {\rm i} \xi (u - U)} =
   \sum_{n=0}^{\infty} \frac{(-1)^n}{n!} \delta^{(n)}(u) U^n.
\end{equation}
Here $u$ is a c--number variable and $U$ an element of the Grassmann--algebra 
built from the matter fields $\psi$ and $\psi^*$. From this definition we get
immediately 
\begin{equation}
\label{delt1}
1 \equiv \int {\rm d}u \, \delta(u-U).
\end{equation} 
Thus, by inserting identites of the form (\ref{delt1}) under the functional 
integral we introduce for each Grassmann--algebra--valued gauge invariant
$(L,R,H,B_{\mu},J_{\mu},J^5_{\mu})$ c--number variables 
$(l,r,h,b_{\mu},j_{\mu},j^5_{\mu})$, which we will call  
c--number mates. These mates are by definition gauge invariant: $r$ and $l$ 
are real scalar fields, $h$ is a complex scalar field and $b_{\mu},j_{\mu}$ 
and $j^5_{\mu}$ are real covector fields. Next we apply the procedure
developed in \cite{KRR1} and \cite{KRR2}: First we use (\ref{3.6}) and 
(\ref{3.7}) under the functional integral to express the Lagrangian in terms 
of gauge invariants. Next we eliminate the auxiliary invariants by the help of 
(\ref{2.7})--(\ref{2.8}) and (\ref{3.2})--(\ref{3.4}). We are left with a 
functional integral, where the original gauge dependent field configuration
$(\psi, A_{\mu})$ occurs only in the integral measure. It can be
integrated out leaving us with the (reduced) functional integral in
terms of gauge invariants (for details we refer to \cite{KRR1} and \cite{KRR2}).
Finally we stress that, due to (\ref{2.7})--(\ref{2.8}), $r$, $l$ and 
$j_{\mu}$ can be expressed by $j^5_{\mu}$. Thus,
for this model we get a reduction of the functional integral (\ref{2.2}) to
an integral in terms of the gauge invariant set
$(v_{\mu},j^{\mu}_5,|h|,\theta)$, where
$h := |h| e^{{\rm i} \, \theta}$ and $v_{\mu} := \frac{b_{\mu}}{e \, |h|^2}$:
\begin{equation}
\label{3.8}
Z = \int \prod_x \left\{ {\rm d}v_{\mu} {\rm d}j^{\mu}_5 \, 
    {\rm d}|h|^2 \, {\rm d}\theta \,
    K[j^{\mu}_5,|h|] \right\} \,
    e^{{\rm i} \, \int {\rm d}^2x \, {\cal L}[v_{\mu},j^{\mu}_5,\theta]},
\end{equation}
with
\begin{eqnarray}
K[j^{\mu}_5,|h|] 
& = & \frac{1}{2 \pi} \, 
      \left( \frac{\delta^2}{\delta j^5_{\mu} \, \delta j_5^{\mu}}
      - \frac{1}{4} \, \frac{\delta^2}{\delta |h|^2} 
      +\frac{1}{4 |h|} \, \frac{\delta}{\delta |h|} \right)
      \delta(j^5_{\mu}) \delta(|h|^2), \label{3.9} \\
{\cal L}[v_{\mu},j^5_{\mu},\theta] 
& = & - \, \tfrac{1}{4} \, \left( \partial_{[\mu} v_{\nu]} \right)^2
      - \, e \, \epsilon^{\mu \nu} \, v_{\mu} \, j^5_{\nu}
      + \tfrac{1}{2} \, j^5_{\mu} \, (\partial^{\mu} \theta). \label{3.10}
\end{eqnarray}	 
We stress that ${\rm d}\theta$ is the Haar--measure on $S^1$.

The effective Lagrangian (\ref{3.10}) leads to the following set of classical 
field equations in terms of gauge invariants:
\begin{eqnarray}
(\partial_{\mu} j^{\mu}_5) & = & 0, \label{3.11} \\
v^{\mu} & = & \tfrac{1}{2 e} \, \epsilon^{\mu \nu} \, (\partial_{\nu} \theta),
   \label{3.12} \\
(\partial_{\mu} \partial^{[\mu} v^{\nu]}) & = & e \, \epsilon^{\nu \mu} \, 
j_{\mu}^5.
   \label{3.13}   
\end{eqnarray}
Formula (\ref{3.11}) means that, on the classical level, the axial
current is conserved. Moreover, from equation (\ref{3.13}) we get
$(\partial_{\mu} j^{\mu}) = 0$,
i.e. also the vector current is conserved.
Finally, equation (\ref{3.12}) leads to 
\begin{equation}
\label{3.15}
(\partial_{\mu} v^{\mu}) = 0.
\end{equation}

We see that, as in the Faddeev--Popov procedure, a nontrivial singular
integral kernel $K[j^{\mu}_5,|h|]$ occurs in the functional integral
(\ref{3.8}). Obviously, the quantity $|h|$ can be integrated out, because it
does not occur in (\ref{3.10}). The remaining kernel 
$K[j^{\mu}_5] = \frac{1}{2 \pi} \, \left( \frac{\delta^2}{\delta j^5_{\mu} \,
\delta j_5^{\mu}} \right) \delta(j^{\mu}_5)$ can be treated similarly as in
the Faddeev--Popov procedure: it can be averaged with a Gaussian measure, or
more generally, with a sum of moments of a Gaussian measure. After this 
``regularization'' a free
parameter, say $\alpha_j$, characterizing the Gaussian measure, enters the
theory. In the Faddeev--Popov approach different values of this parameter
correspond to different gauge fixings. Here, as will be seen in what follows,
this parameter is fixed, e.g. by the physical requirement to give the correct
chiral anomaly. After this ``averaging'' procedure we get
\begin{equation}
\label{3.17}
Z = {\cal N} \int \prod_x \left\{ {\rm d}v_{\mu} {\rm d}j^{\mu}_5 \, 
    {\rm d}\theta \right\} \,
    e^{{\rm i} \, \int {\rm d}^2x \, {\cal L}[v_{\mu},j^{\mu}_5,\theta]},
\end{equation}
where 
\begin{equation}
\label{3.18}
{\cal L}[v_{\mu},j^5_{\mu},\theta] 
 = - \, \tfrac{1}{4} \, \left( \partial_{[\mu} v_{\nu]} \right)^2
   - \, e \, \epsilon^{\mu \nu} \, v_{\mu} \, j^5_{\nu}
   + \tfrac{1}{2} \, j^5_{\mu} \, (\partial^{\mu} \theta)
   - \tfrac{1}{2 \alpha_j} \, j^{\mu}_5 j_{\mu}^5 \, .
\end{equation}
Here, $\cal N$ is an infinite normalization constant,  which
we omit in what follows. We stress that the same procedure works also
for the standard generating functional.

\setcounter{equation}{0}
\section{Standard Bosonisation Rule and Chiral Anomaly}

Since $j_{\mu}^5$ does not occur as a dynamical field a further reduction of 
(\ref{3.17}) is possible, namely we can carry out the Gaussian integration over 
$j_{\mu}^5$. This leads to:
\begin{equation}
\label{4.1}
Z = \int \prod_x \left\{ {\rm d}v_{\mu} \, {\rm d}\theta \right\} \,
    e^{{\rm i} \, \int {\rm d}^2x \, {\cal L}[v_{\mu},\theta]},
\end{equation}
where 
\begin{equation}
\label{4.2}
{\cal L}[v_{\mu},\theta]
 = - \, \tfrac{1}{4} \, \left( \partial_{[\mu} v_{\nu]} \right)^2
   - \tfrac{\alpha_j \, e^2}{2} \, v_{\mu} v^{\mu}
   - \tfrac{\alpha_j \, e}{2} \, \epsilon^{\mu \nu} \, v_{\mu} (\partial_{\nu} 
\theta)
   + \tfrac{\alpha_j}{8} \, (\partial_{\mu} \theta)(\partial^{\mu} \theta).
\end{equation}
After performing the reparameterisation
$\theta(x) \rightarrow 2 \sqrt{\pi} \theta(x)$ as well as fixing the parameter
$\alpha_j = \frac{1}{\pi}$ we obtain:
\begin{equation}
\label{4.3}
{\cal L}
 = - \tfrac{1}{4} \, \left( \partial_{[\mu} v_{\nu]} \right)^2
   - \tfrac{e^2}{2 \pi} \, v_{\mu} v^{\mu}
   - \tfrac{e}{\sqrt{\pi}} \, \epsilon^{\mu \nu} \, {v_{\mu}}
     (\partial_{\nu} \theta)
   + \tfrac{1}{2} \, (\partial_{\mu} \theta)(\partial^{\mu} \theta).
\end{equation}   
This result has to be compared with other functional integral approaches
explaining the bosonisation phenomenon, see \cite{DN1,DN2,DN3}.
Essentially, our procedure leads to the same result, but with
the following advantages: Within our approach, instead of the electromagnetic
potential in a certain gauge, the gauge invariant field $v_{\mu}$ occurs.
It becomes -- automatically -- massive, with mass
$m_v = \frac{e}{\sqrt{\pi}}$. Thus, we see a dynamical Higgs mechanism
characteristic for the Schwinger model, see \cite{AA}, \cite{AB}, \cite{B}
and references therein. Moreover, the field $\theta$, which in other
approaches has to be rather introduced by hand, shows up as the ``phase'' of a
gauge invariant combination of the original fermionic fields. This gives -- in
our opinion -- a deeper insight into the bosonisation phenomenon.
Comparing the couplings of $v_{\mu}$ in (\ref{3.18}) and (\ref{4.3}),
we read off the famous ``bosonisation rule''
\begin{equation}
\label{4.4}
\tfrac{1}{\sqrt{\pi}} \, (\partial^{\mu} \theta) = j^{\mu}_5.
\end{equation}

Next, we show that within our approach the chiral anomaly can be easily
calculated, justifying -- in particular -- the above made choice
$\alpha_j = \frac{1}{\pi}$.
For this purpose, we treat $v_{\mu}$ as an external field. Due to the
bosonisation rule (\ref{4.4}), the chiral anomaly should be given by the
vacuum expectation value
\begin{equation}
\label{4.5}
<\tfrac{1}{\sqrt{\pi}} (\partial^{\mu} \theta)> 
 = \frac{1}{Z[v_{\mu}]} \, \int {\rm d}\theta \,
   \tfrac{1}{\sqrt{\pi}} (\partial^{\mu} \theta) \,
     e^{i \int d^2x {\tilde {\cal L}}},
\end{equation}	
where $Z[v_{\mu}] = \int d\theta \, e^{i \int d^2x {\tilde{\cal L}}}$ and
${\tilde{\cal L}}= -\frac{e^2}{2 \pi} \, v_{\mu} v^{\mu}
- \frac{e}{\sqrt{\pi}} \, \epsilon^{\mu \nu} \, v_{\mu} (\partial_{\nu} \theta)
+ \frac{1}{2} \, (\partial_{\mu} \theta)(\partial^{\mu} \theta)$.
Since $v_{\mu}$ is an external field, i.e. it obeys the classical field
equation (\ref{3.15}), we can write it in the form
$v_{\mu} = \epsilon_{\mu \nu} (\partial^{\nu} \Lambda)$. 
Thus, performing the shift 
$(\partial^{\mu} \theta)
\rightarrow (\partial^{\mu} \theta)
+ \frac{e}{\sqrt{\pi}} \epsilon^{\nu \mu} v_{\nu}
\equiv (\partial^{\mu} \theta)
- \frac{e}{\sqrt{\pi}} (\partial^{\mu} \Lambda)$,
which leads to a trivial change in the integration variable 
$\theta(x) \rightarrow \theta(x)-\frac{e}{\sqrt{\pi}} \Lambda(x)$,
we get:
\begin{eqnarray*}
<\tfrac{1}{\sqrt{\pi}} (\partial^{\mu} \theta)> 
& = & \frac{1}{Z[v_{\mu}]} \, \int {\rm d}\theta \,
      \tfrac{1}{\sqrt{\pi}} \left( (\partial^{\mu} \theta) 
      + \tfrac{e}{\sqrt{\pi}} \epsilon^{\nu \mu} v_{\nu} \right)
      e^{{\rm i} \int {\rm d}^2x \{ \frac{1}{2} \, 
	(\partial_{\mu} \theta)(\partial^{\mu} \theta) \} } \\
& = & \frac{1}{Z[v_{\mu}]} \, \int {\rm d}\theta \,
      \tfrac{1}{\sqrt{\pi}} (\partial^{\mu} \theta) 
      e^{{\rm i} \int {\rm d}^2x \{ \frac{1}{2} \, 
	(\partial_{\mu} \theta)(\partial^{\mu} \theta) \} } 
    + \tfrac{e}{\pi} \epsilon^{\nu \mu} v_{\nu}.
\end{eqnarray*}  
Now observe that the first integral is odd in $\theta$ and, consequently, it
vanishes identically, leading to
\begin{equation}
\label{4.6}
<\tfrac{1}{\sqrt{\pi}} (\partial^{\mu} \theta)>
 = \tfrac{e}{\pi} \epsilon^{\nu \mu} v_{\nu}.
\end{equation}
Taking the partial derivative gives exactly the chiral anomaly:
\begin{equation}
\label{4.7}
\partial_{\mu} <\tfrac{1}{\sqrt{\pi}} (\partial^{\mu} \theta)>
 \equiv <\tfrac{1}{\sqrt{\pi}} (\partial_{\mu} \partial^{\mu} \theta)>
    =	\tfrac{e}{\pi} \epsilon^{\nu \mu} (\partial_{\mu} v_{\nu}).
\end{equation}
We see that within our formulation the chiral anomaly can be calculated in a
straightforward way without using neither perturbative techniques nor
regularization techniques of the heat kernel type. We also stress that this
calculation can be -- a posteriori -- considered as a deeper justification of
the bosonisation rule (\ref{4.4}).

Moreover, we have
\begin{equation}
\label{4.8}
<\tfrac{1}{\sqrt{\pi}} \epsilon^{\nu \mu} (\partial_{\nu} \theta)>
 = \tfrac{e}{\pi} v^{\mu}.
\end{equation}
Using again the classical field equation (\ref{3.15}) yields
\begin{equation}
\label{4.9}
\partial_{\mu} <\tfrac{1}{\sqrt{\pi}} \epsilon^{\nu \mu}
(\partial_{\nu} \theta)>
 = \tfrac{e}{\pi} (\partial_{\mu} v^{\mu})
 \equiv 0.
\end{equation}
Due to (\ref{4.4}) and the relation between vector and axial current,
equation (\ref{4.9}) shows the conservation of the vector current $j_{\mu}$.

Finally, let us make some remarks. If we consider the massive Schwinger
model, where an additional term $-m \overline{\psi} \psi$ occurs in the
Lagrangian (\ref{2.3}), we get an additional term
$-2 m {\rm Re}(h) = -2 m |h| {\cos \theta}$ in (\ref{3.3}). This leads to a
field theory of sine--Gordon type. Comparing with the standard result,
see for instance \cite{C} and \cite{Fr}, we obtain a slight
modification: instead of getting a constant in front of
the $\cos(\theta)$--term, we obtain $|h|$. However, since $|h|$ is a
non-propagating field, it can be ``averaged'' to a constant -- along the
same lines as discussed at the end of Section 3.

Our approach also works for the 2--dimensional gauged Thirring--model
(see for instance \cite{SW}, where the Thirring--model and its various 
generalizations on curved space--time are analyzed).
Applying our procedure simply leads to a modification of the
coefficient of the term quadratic in $j_{\mu}^5$ occuring in equation
(\ref{3.18}).

Similar investigations for full QED are in progress.

\section*{Acknoledgement}

The authors are very much indebted to I. Bialynicki-Birula for helpful
discussions and remarks.


\end{document}